\documentclass[aps,prl,twocolumn,superscriptaddress,groupedaddress]{revtex4}
\usepackage{subfig}
\usepackage{graphicx}  
\usepackage{dcolumn}   
\usepackage{bm}        
\usepackage{amssymb}   
\usepackage{slashed}
\usepackage{graphicx}				
\usepackage{amsmath}
\usepackage{mathtools}
\usepackage{tikz}
\usepackage{tikz-cd}
\usetikzlibrary{arrows,backgrounds}
\usepackage[all]{xy}
\usepackage{yfonts}
\usepackage{slashed}
\usepackage{color}

\begin{document}
\title{Global aspects of the renormalization group and the Hierarchy problem}
\author{Andrei T. Patrascu}
\address{University College London, Department of Physics and Astronomy, London, WC1E 6BT, UK}

\begin{abstract}
The discovery of the Higgs boson by the ATLAS and CMS collaborations allowed us to precisely determine its mass being 125.09 $\pm$ 0.24GeV. This value is intriguing as it lies at the frontier between the regions of stability and meta-stability of the standard model vacuum. It is known that the hierarchy problem can be interpreted in terms of the near criticality between the two phases. The coefficient of the Higgs bilinear in the scalar potential, $m^{2}$, is pushed by quantum corrections away from zero, towards the extremes of the interval $[-M^{2}_{Pl},M^{2}_{Pl}]$ where $M_{Pl}$ is the Planck mass. In this article, I show that demanding topological invariance for the renormalisation group allows us to extend the beta functions such that the particular value of the Higgs mass parameter observed in our universe regains naturalness. In holographic terms, invariance to changes of topology in the bulk is dual to a natural large hierarchy in the boundary quantum field theory. The demand of invariance to topology changes in the bulk appears to be strongly tied to the invariance of string theory to T-duality in the presence of H-fluxes. 
\end{abstract}
\maketitle
\section{Introduction}
The discovery of the Higgs boson and the measurement of its mass as lying at the very edge between EW stability and metastability regions, not accompanied by the discovery of supersymmetric particles, already casts shadows of doubt upon a potential supersymmetric solution to the hierarchy problem. This state of affairs asks for the search of new principles of nature that may allow us to understand the origins of the standard model parameters without having to rely on symmetries and without having to give up calculability as is the case with multiverse interpretations. As recent experimental results suggest, the standard model vacuum appears to be metastable. Understanding the parameters of the standard model that lead to such a conclusion would therefore reveal information about the future fate of our universe. In this article, I show that naturalness can be restored if one imposes invariance of the physical theory with respect to topology changes associated to the renormalisation group. 
Let $BG$ be the classifying space of the (in this case renormalisation) group $G$ constructed as the quotient $EG/G$, where $EG$ is a contractible space on which $G$ acts freely.

By means of the isomorphism in cohomology 
\begin{equation}
H^{n}(BG,\mathbb{Z})=H^{n}(G,\mathbb{Z})
\end{equation}
it is clear that the topological properties (as described by cohomology) of the renormalisation group $G$, are equivalent to the topological properties of the space on which $G$ acts. According to the duality between renormalisation group equations of the low energy theory and the evolution equations along the new directions in a higher dimensional effective theory (e.g. 5-d Supergravity) [1-3] the cohomology of the renormalisation group is to be associated to the cohomology of the subspace induced by the evolution equations along the additional directions in a higher energy effective theory. Therefore, imposing invariance with respect to the change in the renormalisation group topology is equivalent to imposing invariance with respect to topology changes of the high energy effective theory. This prescription is the low energy "remnant" of the string theoretical T-duality.

Therefore, understanding how the principle of topological invariance, observed initially in string theory, affects low energy effective field theories like the standard model or 5-d supergravity could herald new ways of connecting low energy physics to the physics determined by extended fundamental objects expected to arise at or near the Planck scale. While the observation that the string theoretical T-duality may be related to the hierarchy problem is not new [1], the fact that the topological invariance implied by T-duality is the key to restoring naturalness, is.

The beta functions (generators of the renormalisation group) allow us to probe high energies starting from our low energy effective field theory. However, at some point they must enter the regime where string geometry effects become relevant. In order to connect lower energy scales to string geometry effects involving topology changing dualities, such beta functions must not discern topologies connected by means of dualities. This requirement alters the topological structure of the renormalisation group such that it compensates for the topology changes that may occur at the string level. The construction of topology invariant theories with a focus on group quantisation was performed in [4]. The methods introduced there are particularly relevant because they deal with group topology.

The creation of explicitly T-duality invariant string theories resolved the questions regarding the high energy domain. Representative articles are [27], [33]. Transferring such invariance to the renormalisation group is the focus of this work. Due to the duality between RG flux and extra dimensional equations of motion, the goal of this article is to explore the field theoretical, boundary effects of the invariance of the theory to topology changes in the bulk. 
Of course, after the methodology is established, new insights upon the cosmological constant, the vacuum stability, or the nature of dark matter may be obtained.  I will only discuss here applications to the hierarchy problem, leaving the other subjects to a future, cosmology oriented article. Of course the standard model hierarchy problem and the cosmological constant problem are most likely related.

In quantum field theory, energy scales are regarded as linearly ordered, from the lowest to the highest. Moving from high energy scales to low energy scales leads to decoupling of the heavy states and to effective theories. The domain where such quantum field theories are valid is also where the renormalisation group is useful, allowing us to move between different energy scales. As we move towards higher energies, in supergravity, we may identify renormalisation group flows with equations of motion on the additional directions. In string theory however, distinguishing the scales is not perfectly defined. T-duality itself must arise by means of a "conspiracy" between physics at all scales simultaneously [5]. T-duality in string theory, together with Buscher's rules for the target space [6, 7], is to be associated with the low energy demand for invariance of physics under a change of topology of the renormalisation group. The plan of this article is as follows. First, I introduce the renormalisation group from an algebraic perspective, connecting the form of the generators (beta functions) with the global topological structure of the group. Next, I impose the invariance of physical results with respect to group topology changes by suitably altering the group laws. In order to take this new principle into account a modification of the beta function will be required. Finally, I will show on a simple quantum field theoretical model how these modifications explain the naturalness without the need of additional symmetries or particles. Of course, the presence of new symmetries at higher scales is not excluded by these results. Supersymmetry may still be a symmetry of nature at higher energies but as experiments suggest, it may not be able to explain naturalness on its own.

\section{Bulk space and the Renormalisation Group}
Given holography and, particularly, the AdS/CFT duality, quantum gravity can be written in terms of a non-gravitational quantum field theory on a boundary, allowing the application of Wilson's renormalisation group. In this context, T-duality is seen as a duality in the bulk space involving string theory. Invariance with respect to T-duality has been made explicit at this level, for example by using a type of space that looks locally like a Riemannian manifold but which is glued together from these local patches not just by diffeomorphisms but also by T-duality transformations along some torus fibres (T-folds). Such a T-fold would be a target space for a string sigma-model that is only locally a Riemannian manifold but globally would look like a more general geometry. T-folds have a description in terms of spaces that locally are fibre products of a torus fibre bundle with its T-dual [34]. Among the first to create an explicitly T-duality invariant formulation for the string world sheet action was Tseytlin [27]. Here, I show that by holographic duality, a demand for topological invariance resulting from T-duality on the bulk, results in the restoration of naturalness in the hierarchy of the boundary quantum field theory.

The gauge/gravity duality can be interpreted, according to [20] as a geometrisation of the quantum field theoretical renormalisation group. While the connection between a higher dimensional equation of motion and the lower dimensional renormalisation group flow was known already in [1], it was [20] who reformulated this connection in modern, holographic terms. The exact renormalisation group equations can be expressed as Hamilton equations for the radial evolution in a model space-time with one higher dimension. 

The jet-bundle structure corresponding to the quantum field theory is employed by [20] in order to show how the geometry corresponding to the RG equations emerges. These equations are being interpreted as higher spin equations of motion. In this article I emphasise mostly the topological aspects of the RG and how dualities connecting topologically distinct string geometries manifest themselves at lower energies as invariances to topological changes of the renormalisation group. To understand such a topological approach to the bulk-space structure, it is important to understand how the generators of the renormalisation group, the beta functions, emerge as geometrical objects. The scale transformations in the field theory correspond to movement in the additional radial dimension [20] and the renormalisation group trajectories are associated to specific geometries. If the RG flow begins or ends near a fixed point, the corresponding geometries are anti-de Sitter and connections between gauge/gravity dualities and RG flows are known [21].

The picture that emerges is that we should formulate the holographic duals of exact RG equations in terms of connections and sections of certain bundles over a model space-time. Indeed, once such a formulation has been constructed the next step will be to establish the topological invariance at low energies that results from the string theoretical T-duality.
Following ref. [20] in the first step of the Wilsonian RG process, we integrate out a shell of fast modes and analyse the effect on the sources. Lowering $M\rightarrow \lambda M$, where $\lambda=1-\epsilon$ is interpreted as an integration of the fast modes, we obtain the variation of the sources 
\begin{equation}
\begin{array}{c}
\delta_{\epsilon}\mathcal{B}=-\epsilon M\frac{d}{dM}\mathcal{B}\\
\delta U=-\epsilon M \frac{d}{dM} U\\
\end{array}
\end{equation}
We demand that for the overall partition function 
\begin{widetext}
\begin{equation}
M\frac{M}{dM}Z=Z_{0}^{-1}\int[d\phi d\phi^{*}]\{(M\frac{d}{dM}e^{iS_{0}}e^{iS_{1}}+e^{iS_{0}}(M\frac{M}{dM}e^{iS_{1}})-Z_{0}^{-1}e^{iS_{0}+iS_{1}}M\frac{d}{dM}Z_{0}\}=0
\end{equation}
\end{widetext}
where the action has been regulated and split into $S=S_{0}+S_{1}$ with $S_{1}$ containing the $\mathcal{B}(x,y)$ sources. 
Following the calculations of [20] we obtain 
\begin{equation}
\begin{array}{c}
\delta_{\epsilon}\mathcal{B}=-\epsilon M\frac{d}{dM}\mathcal{B}=\epsilon\mathcal{B}\Delta_{B}\mathcal{B}\\
\delta_{\epsilon}U=-\epsilon M \frac{d}{dM}U=-i\epsilon N Tr(\Delta_{B})\mathcal{B}\\
\end{array}
\end{equation}
The second step of the renormalisation group approach is to perform a transformation that brings the cutoff back while changing the conformal factor of the metric
\begin{equation}
\mathcal{L}(x,y)=\delta^{d}(x,y)+\epsilon z W_{z}^{(0)}(x,y)
\end{equation}
and hence we obtain, again, making use of [20]
\begin{equation}
\begin{array}{c}
\mathcal{B}(z+\epsilon z)=\mathcal{B}(z)-\epsilon[W_{z}^{(0)},\mathcal{B}]+\epsilon \mathcal{B}\Delta_{B}\mathcal{B}\\
U(z+\epsilon z)=U(z)-i\epsilon N Tr(\Delta_{B})\mathcal{B}
\end{array}
\end{equation}
We may redefine 
\begin{equation}
\Delta_{B}=\frac{M}{z}\frac{d}{dM}(D^{2}_{(0)})^{-1}
\end{equation}
and this allows us to extend the definition of the sources to the whole of the RG bulk space. 


The connection between the holographic renormalisation group in the bulk and the Wilsonian renormalisation group in the dual field theory on the boundary has been discussed in [30] where the Wilson RG transformation is mapped by the AdS/CFT duality and takes a holographic form.

At the string scale the formalism works as for the T-duality invariant doubled string formalism. The UV-divergences in the doubled formalism were studied via background field expansion in [39]. Their effects on the beta functions have also been studied and it was noticed that in the doubled interpretation, the high energy beta functionals are to be analysed in a target space that has been dimensionally reduced so that only the base coordinates remain [40]. This shows that the high energy beta functions have been made topologically invariant by means of the doubling string interpretation, and there are no additional UV instabilities produced at that level. The IR region can be reached by changing the coefficient structure in cohomology from one allowing extended objects to one allowing only points. Demanding that these two situations are dual according to the universal coefficient theorem and hence demanding for the associated groups to be isomorphic both as original groups and as extensions leads to an additional co-boundary term in the group law. Such a term represents a unification of low and high energy scales and, according to the universal coefficient theorem, by using the $Ext$ group, leads to a unified interpretation of both IR and UV domains. Therefore this cobordism stabilises also the IR region (see next section). 


Scale transformations in the field theory correspond to movement in the extra dimension and therefore the specific RG trajectories correspond to particular bulk geometries. On the quantum field theoretical side, the RG transformations in a perturbative context are regarded as deformations away from a free RG fixed point [20]. In holography, simple geometric constructions in the bulk correspond to strongly coupled dynamics in the dual field theory and hence the free fixed point RG discussion on the field theoretical side may seem incompatible with the standard holographic image. This point of view is however too superficial. In analysing the conjectured duality between free vector models in $d=2+1$ dimensions and higher spin theories on $AdS_{4}$ it was noted in [31] and [32] that the field theory side is well understood while the bulk theory was highly non-linear, containing arbitrary high spins. An example used in [20] was the connection between three dimensional Chern-Simons theories known to be dual to two dimensional Wess-Zumino-Witten models. The theory in the bulk is now topological, meaning that it does not depend on the bulk metric and diffeomorphism invariance is broken only on the boundary by terms explicitly containing the boundary metric [20]. From studying the exact RG equations, [20] shows that the holographic duals of vector models can be expressed in terms of connections and sections of certain bundles over a model spacetime. The goal of this paper is to further generalise this to a situation where topology transformations in the bulk, induced by the action of a T-duality transformation, are supposed to provide the same physics. In this case, insensibility to bulk topology changes is understood by means of the methods discussed in [20] as modifications of the beta functions in the field theory sector that are capable of restoring naturalness of the hierarchy. Otherwise stated, invariance to topology changes in the bulk is dual to a natural hierarchy in the quantum field theory. Such topology changes are of the type induced by T-duality in the presence of non-trivial H-fluxes. Of course, imposing topological invariance is already a non-perturbative criterion imposed on the bulk, hence the method has already from the construction non-perturbative stability.

When we parametrise the infinitesimal RG transformation according to [20] by writing it as 
\begin{equation}
\mathcal{L}=1+\epsilon\cdot z\cdot W_{z}^{(0)}
\end{equation}
the second step changes $z\rightarrow \lambda^{-1}z$ and hence the RG flow becomes parametrised by $z$ instead of $M$ which becomes just an auxiliary parameter in the cutoff function. Here, $z$ represents a conformal factor in the background metric $\eta_{\mu\nu}\rightarrow z^{-2}\eta_{\mu\nu}$ while the sources $B$ and $W$ are transformed as $B_{old}=z^{d+2}B_{new}$ and $W_{old}=z^{d}W_{new}$. The effective renormalisation scale will therefore be $\mu=\frac{M}{z}$ and the renormalisation group flow will be parametrised by $z$. This will then be interpreted as a bulk coordinate. The infinitesimal piece of $\mathcal{L}$ is $W_{z}^{(0)}$ which can be thought as the $z$ component of the connection. $W^{(0)}_{z}$ is regarded as a bookkeeping device, keeping track of the gauge transformations along the RG flow. In order to use the notation of [20] one may reabsorb the tensorial components of $W$ into $B$ and to redefine $B$ as $\mathcal{B}=B-\{\hat{W}^{\mu},D_{\mu}^{(0)}\}-\hat{W}_{\mu}\cdot\hat{W}^{\mu}$ where $D_{\mu}^{0}$ is the covariant derivative defined in [20]. In this context, where a free bosonic vector model is considered to be perturbed away from the fixed point by the bi-local source $\mathcal{B}$, and where the fields $\mathcal{B}$ and $W^{(0)}$ are extended into the entire RG mapping space (the RG bulk space), we may interpret and re-organise the renormalisation group equations as covariant equations with the beta function playing the role of a curvature. Let the bulk extension of $\mathcal{B}$ be called $\textfrak{B}$ and the bulk extension of $W^{(0)}$ be called $\mathcal{W}^{(0)}$. In ref. [20] it is shown that the renormalisation group equations emerge as gauge-covariant equations in the bulk. With $\mathcal{W}_{\mu}^{(0)}$ flat in the transverse directions by construction we can write 
\begin{equation}
\mathcal{F}^{(0)}=\textbf{d}\mathcal{W}^{(0)}+\mathcal{W}^{(0)}\wedge \mathcal{W}^{(0)}=0
\end{equation}
where $\textbf{d}=dx^{\mu}\cdot [\partial_{\mu},\cdot]+dz\cdot\partial_{z}\cdot$ is the bulk exterior derivative. This being a boundary operator it satisfies $\textbf{d}^{2}=0$. For the extension of the $\mathcal{B}$ field we have 
\begin{equation}
\partial_{z}\textfrak{B}+[\mathcal{W}_{z}^{(0)},\textfrak{B}]=\beta^{(\textfrak{B})}
\end{equation}
The $z$ component of the $1$-form $\beta^{(\textfrak{B})}=\beta_{\mu}^{(\textfrak{B})}dx^{\mu}+\beta^{(\textfrak{B})}dz$ is given by 
\begin{equation}
\beta^{(\mathcal{B})}=\mathcal{B}\cdot \Delta_{B}\cdot\mathcal{B}
\end{equation}
with $\Delta_{B}=\frac{M}{z}\frac{d}{dM}(D_{\mu}^{(0)-2})$ where $D_{\mu}^{(0)-2}$ is the covariant derivative for a background gauge symmetry as defined in [20]. The beta function is therefore interpreted as a curvature. A series of topological invariants arise as integrals over curvatures. It is interesting to note that in the context of [22] and [23], there is no measurement in general relativity that can unambiguously detect the presence of a generic wormhole geometry. This statement is equivalent, due to ER-EPR with the statement that entanglement is not detectable. Such an observation is in agreement with the fact that (co)homology with different choices of coefficients may detect certain aspects of a topological space while masking others and therefore, that physics must remain invariant with respect to topology changes, due for example to string theoretical T-duality transformations. Now that a geometrical interpretation of the beta function has been established [20], it remains to be seen how the invariance with respect to topology changes affects the beta functions. Many topological invariants can be expressed in terms of integrals over various differential geometrical functions like curvature, Gauss curvature, etc. In order to make the low energy results invariant to topological changes, the integrals over the beta functions must not change when topological changes due to T-duality are invoked. To see how this works it is important to first consider the string theoretical duality symmetry.

\section{Topological aspects of the Renormalisation Group}
Let $g(l)$ be the renormalisation group transformation such that, when it acts on a parameter of the theory it obeys the group law $g(l)\times g(\lambda)=g(l+\lambda)$. Given a parameter of the theory $m$, we have $g(l)m=m'=G(l,m)$. $G$ is a continuous function of the two parameters satisfying the normalisation condition $G(0,m)=m$ corresponding to the transformation $g(0)=Id$. We can of course transform the group law from this algebraic form to a functional form writing $G(l,G(\lambda, m))=G(l+\lambda,m)$, and in the infinitesimal form we obtain the well known expression $\frac{\partial G(l,m)}{\partial l}=\beta (G(l,m))$. We can clearly see that the group generator is $\beta(m)=\frac{\partial G(\epsilon,m)}{\partial \epsilon}$ at $\epsilon = 0$. Let there be a physical quantity $F(Q^{2},m)$ calculated under a certain renormalisation prescription as $F(\frac{Q^{2}}{\mu^{2}},m_{\mu})$, $m_{\mu}$ being the renormalised parameter (coupling) at some renormalisation point $Q=\mu$. Demanding that $F$ does not depend on the energy scale reads $\frac{dF}{d\mu}=0$, or, making use of the parameters on which $F$ depends 
\begin{equation}
[x\frac{\partial}{\partial x}-\beta(m)\frac{\partial}{\partial m}]F(x,m)=0
\end{equation}
where $x=\frac{Q^{2}}{\mu^{2}}$, and $m=m_{\mu}=\bar{m}(\mu^{2})$ is called the effective coupling, $\beta$ being the group generator defined above. As they stay, the renormalisation group equations represent simply the group laws and do not contain any physics. The change in the parameters of the theory and the energy scale where the measurements are performed do obey such a group law. However, when the energy scale becomes high enough for string geometry to play a role, we need to take into account the possibility of topology change and to make the observable results independent of such a change. One way in which this can be taken into account is by means of cohomology with non-trivial coefficients and the use of the universal coefficient theorem. 

Non-trivial coefficients in cohomology represent the topological and geometrical structure associated to a point of the manifold upon which the group acts. When the coefficients are trivial the point has no additional structure. When the coefficients are non-trivial the point may get the structure of various extended objects (strings, branes). The fact that the coefficient structure in cohomology encodes the algebraic structure added to a point on the manifold, giving it the properties of extended objects has been discussed in [38]. In this sense, the universal coefficient theorem, by allowing us to move from one coefficient structure to another, can allow us to move from the trivial structure of a point (classical geometry) to an extended structure with various topologies (stringy geometry). This property is of utmost importance. Properly analysed, it may reveal the origin of the holographic principle and of AdS/CFT. Otherwise stated, this exact sequence in homological algebra potentially encodes several classes of dualities between string theoretical constructions and field theoretical ones. Locality on the boundary is associated to coefficients in cohomology encoding ordinary points. Universal coefficient theorems may define exact sequences connecting such local quantum field theories with non-local theories in the bulk, associated to coefficients in cohomology encoding "generalised" points i.e. algebraic curves, bundles, etc. Understanding the maps arising in the universal coefficient theorem may therefore reveal new dualities and new ways in which certain structures may be seen from a dual perspective. Using the universal coefficient theorem to define new dualities, beyond holography or the ER-EPR conjecture is the subject of future articles. Here, I will use it to reinterpret the hierarchy problem from a different perspective. 
While a classical point is restricted in probing the topology of a manifold, a string will be able to detect non-trivial topology due to its ability of wrapping around a non-trivial cycle. There is however another way of probing topology, even at lower energies, by means of quantum field theories defined via path integrals. Indeed quantum field theories integrate all available paths which therefore will have access to the global structure of the manifold. A discussion on this aspect of path integral quantisation may be found in [41]. 

Suppose we have already reached the scale where extended structures become relevant. For group cohomology, the universal coefficient theorem is 
\begin{widetext}
\begin{equation}
0\rightarrow Ext(H_{p-1}(G;M_{1}),M_{2})\rightarrow H^{p}(G;M_{2})\xrightarrow{h} Hom(H_{p}(G;M_{1}),M_{2})\rightarrow 0
\end{equation}
\end{widetext}
where $G$ is our group, $H_{q}$ represents the $q$-th order homology while $H^{q}$ represents the $q$-th order cohomology, and $M_{1}$, resp. $M_{2}$ represent the coefficients that make the trivial resp. non-trivial topology of the target space manifest. 
The sequence above is exact. Were it not for the $Ext$ group on the left, the arrow $h$ would have been an isomorphism. Clearly, the $Ext$ group is what will signal to our beta function the distinction given by different topologies. In order to make our renormalisation group equation insensible to changes in topology and therefore to be able to take into account $T$-duality with non-trivial $H$-fluxes I will show what group laws are permitted. Let by notation call $H_{p-1}(G,M_{1})=G_{1}$. Then we are interested in $Ext(G_{1},M_{2})$. $G_{1}$ controls the topological aspects of our renormalisation group $G$ as seen by means of homology with coefficients in $M_{1}$. The extension will henceforth be called $\tilde{G}$. Its group elements $\tilde{g}\in\tilde{G}$ are $\tilde{g}=(m,g)$, with $m\in M_{2}$. The group law of $\tilde{G}$ is 

\begin{equation}
(m',g')(m,g)=(m'+m+\xi(g',g),g'g)
\end{equation}
Let there be two extensions $\tilde{G}_{1}$ and $\tilde{G}_{2}$ associated to two potential universal coefficient sequences. Starting for example with the trivial coefficient structure $\mathbb{Z}$ which would encode ordinary points and therefore local quantum field theory, the two universal coefficient theorems would expose the maps that would lead to target manifolds of different topologies (for example spheric and toroidal). 
Let them be
\begin{equation}
\begin{array}{c}
1\rightarrow K \xrightarrow{i_{1}}\tilde{G}_{1}\xrightarrow{\pi_{1}} G\rightarrow 1\\
\\
1\rightarrow K \xrightarrow{i_{2}}\tilde{G}_{2}\xrightarrow{\pi_{2}} G\rightarrow 1\\
\end{array}
\end{equation}

with the two extensions related by a map $\tilde{f}=\tilde{G}_{1}\rightarrow \tilde{G}_{2}$ such that $i_{2}=\tilde{f}\circ i_{1}$ and $\pi_{1}=\pi_{2}\circ \tilde{f}$. We have therefore two group laws, associated to two different two-cocycles defining the group extensions $\tilde{G}_{1}$ and $\tilde{G}_{2}$. I will distinguish the two by using round brackets for the first group law and square brackets for the second group law (basically, using the notation and derivation given in [8])
\begin{widetext}
\begin{equation}
\begin{array}{cc}
(m',g')(m,g)=(m'+m+\xi_{1}(g',g),g'g), & [m',g'][m,g]=[m'+m+\xi_{2}(g',g),g'g]\\
\end{array}
\end{equation}
\end{widetext}
If we assume there exists an isomorphism $\tilde{f}:\tilde{G}_{1}\rightarrow \tilde{G}_{2}$ then, since $(m,g)=(m,e)(0,g)$ and since $\tilde{f}$ is a homomorphism, it is clear that $\tilde{f}$ is fully determined once the images of the elements $(m,e)$ and $(0,g)$ are given. But then a commutative diagram arises when the two exact sequences above are glued together at the left and right ends and the isomorphism $\tilde{f}$ is employed to connect the two extensions in the middle. For the diagram to be commutative, a set of conditions will appear on $\tilde{f}$
\begin{equation}
\begin{array}{c}
\tilde{f}\circ i_{1}=i_{2} \; \Rightarrow \; \tilde{f}(m,e)=[m,e]\\
\\
\pi_{2}\circ \tilde{f}=\pi \; \Rightarrow \; \tilde{f}(0,g)=[\eta(g),g]\\
\end{array}
\end{equation}
This implies that $\tilde{f}$ must have a generic form $\tilde{f}(m,g)=[m+\eta(g),g]$. When we know $\eta(g)$ we have defined $\tilde{f}$. But we do know that $\tilde{f}$ is a homomorphism, hence
\begin{equation}
\tilde{f}(m'+m+\xi_{1}(g',g),g'g)=[m'+m+\xi_{1}(g',g)+\eta(g'g),g'g]
\end{equation}
must be equal to 
\begin{equation}
\begin{array}{c}
\tilde{f}(m',g')\tilde{f}(m,g)=[m'+\eta(g'),g'][m+\eta(g),g]=\\
=[m'+m+\xi_{2}(g',g)+\eta(g')+\eta(g),g'g]\\
\end{array}
\end{equation}
and hence $\eta(g)$ must satisfy the equality 
\begin{equation}
\begin{array}{l}
\xi_{1}(g',g)=\xi_{2}(g',g)+\eta(g')+\eta(g)-\eta(g'g)=\\
=\xi_{2}(g',g)+\xi_{cob}(g',g)\\
\end{array}
\end{equation}
where $\xi_{cob}(g',g)$ is the two-coboundary generated by $\eta(g)$. Therefore, when the diagram is commutative $\xi_{1}$ and $\xi_{2}$ define the same extension. 
When $\xi_{1}$ and $\xi_{2}$ are proportional i.e. $\xi_{2}=\lambda \xi_{1}$ then the last equality cannot be satisfied for all $g'$ and $g$ and therefore they define isomorphic groups $\tilde{G}_{1}$ and $\tilde{G}_{2}$ that are different as extensions. Otherwise stated, they define different elements of the second cohomology group associated to the original group.

\begin{figure}
\centering
\includegraphics[width=130pt]{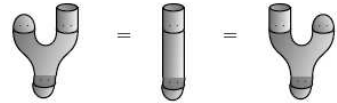}
\caption{The cobordism modifying the group operation connects two distinct high energy topologies, making them indistinguishable in the low energy theory. The doubling method assures indistinguishability in the explicitly T-dual formulation of string theory. The lower part of the figures represent the low energy, the upper part represents string theoretical topology changing effects.}
\end{figure}

This implies that for the group laws $(m',g')(m,g)=(m'+m+\xi_{1}(g',g),g'g)$, the correction to a simple addition $m'+m$, namely $\xi_{1}$, must satisfy the equality $\xi_{1}(g',g)=\xi_{2}(g',g)+\eta(g')+\eta(g)-\eta(g'g)=\xi_{2}(g',g)+\xi_{cob}(g',g)$. This translates into an additional freedom in defining the beta functions and the introduction of additional terms. 

In the context of the renormalisation group, the group laws defined above are basically the renormalisation group equations governing the flows of the parameters of an effective field theory. The demand that the renormalisation group equations and the respective flows be independent of the topology changes emerging from stringy features at high energies, results in the requirement that the groups obtained above are not distinguishable as extensions. In figure 1 it is explained how different topologies in the bulk (the upper level), when connected by means of a cobordism may lead to boundary effective field theories (lower level in the diagrams of Figure 1) that would not distinguish the underlying topological structure. While the string theoretical topologies are connected by T-duality, the freedom of adding a cobordism in the low energy beta function results in a form of invariance to renormalisation group changes in topology.

\section{T-duality}
As a generalisation of the $R\rightarrow \frac{1}{R}$ invariance of string theory compactified on a circle of radius R, T-duality transformations in the case of the low energy effective field theory are given by the Buscher rules [6-7]. However, the application of such rules is restricted in the topologically non-trivial case due to their validity only over local spacetime patches. In the case of the topologically non-trivial Neveu-Schwarz 3-form H-flux, several T-duals have been found and it has been noticed that T-duality not only changes the H-flux but also the spacetime topology [9-11]. It is known that low energy supersymmetry solves the hierarchy problem elegantly. However, from the point of view of the low energy effective actions it may happen that the original and T-dual theories do not have the same number of spacetime supersymmetries [12]. In certain extreme cases, supersymmetry may completely disappear after performing a T-duality operation. We may look at the T-duality transformation as to a change of variables and therefore the symmetries of the original theory are expected to be preserved, although when those symmetries do not commute with the duality, they may be realised only non-locally. That perturbative string theoretical calculations may not detect supersymmetry which is however visible in M-theory is a well known fact that led the authors of [13] to discuss the so called phenomenon of "supersymmetry without supersymmetry". Strings, unlike point particles, probe the target spacetime differently, leading to various effects that may not have analogues in theories based on dimensionless objects. Among these effects, the non-localisation of the extended worldsheet and associated target space  supersymmetry under a T-duality transformation appears to be particularly important. A given conformal field theory may appear differently when analysed from different target space perspectives related by T-duality. If we formulate T-duality transformations in the form of canonical transformations of the worldsheet theory, the coordinate with respect to which duality is performed (say $\eta$) and the corresponding coordinate in the dual theory (say $\tilde{\eta}$) are non-local functions of each other. When we integrate over the string length parameter which appears in the relation between $\eta$ and $\tilde{\eta}$ the non-locality emerges. Any $\eta$ dependent quantity in one theory becomes a non-local function of the corresponding coordinate in the dual theory. To this effect one adds also the interchange between momentum modes (local) and winding modes (non-local) which is also exclusively due to the extended nature of the string. When the worldsheet theory has an extended supersymmetry, if a complex structure associated to such an extended supersymmetry does not have a dependence on the coordinate $\eta$, then in the dual model the extended supersymmetry is realised in the usual way. However, there exist situations when supersymmetry is not preserved by T-duality as noticed above. In all these situations, the complex structure associated with the supersymmetry depends on the coordinate $\eta$ and in the dual theory the complex structure is replaced by a non-local object. In these cases, the extended supersymmetry in the dual theory, albeit still present, is realised only non-locally. The relation between supersymmetry and target space geometry is modified [14]. 
However, non-local effects may be eliminated by changes in the underlying structure associated via coefficients in cohomology to every point of the space. Such modifications are allowed exclusively by freedoms given by the stringy nature of the high energy domain but in the low energy domain they allow for corrections in the beta functions and the RG flow equations identified as transformation group laws. 
\section{Duality symmetric string world sheets}
Having explained how the renormalisation group flows can be interpreted in geometric terms and, following [20], how the beta function can be interpreted as a geometric curvature, it remains to be seen how string theory can be formulated in a T-duality symmetric way so that we can easily transfer its topological invariance to low energy effective theories. String theory implies the existence of a target space in which the dynamics of strings manifests itself. This target space is usually associated with spacetime and allows the construction of S-matrix like observables that connect with low energy local effective field theories.  However, the idea that spacetime itself may be emergent and that the dynamics and interaction of strings may determine its geometrical properties seems to be gaining some momentum [25], [26]. Moreover, at low energies, string theory possesses duality symmetries which relate apparently different supergravity backgrounds. The string spectrum is known to contain winding modes which are capable of probing spacetime in a different way. T-duality is defined in terms of exchanging the winding and momentum modes while changing the gravity background to render physics invariant. The mere existence of T-duality shows that strings experience geometry differently compared to point particles. It therefore makes sense to keep both the winding and momentum modes on equal footing. This has been done [27], [28] by reformulating string theory in a manifest duality symmetric way. As a consequence the number of dimensions of the target spacetime doubled. The extended nature of such a target space allowed the symmetric inclusion of both a geometry and its T-dual. The background equations of motion for the doubled target space are given by the so called Double Field Theory (DFT). This realises T-duality as a manifest symmetry and incorporates both the momentum modes and the winding modes [27]. This method doubles the spacetime dimension with the doubled dimensions being related by T-duality. Additional degrees of freedom will then appear which will have to be reduced by means of chirality constraints. From a spacetime perspective T-duality is a solution generating symmetry of the low energy equations of motion. From a world sheet point of view, T-duality is a non-perturbative symmetry [36]. The presence of T-duality allows for the construction of non-geometric manifolds where locally geometric regions are patched together by means of T-duality transformations. An analysis of the beta functional of the string in the doubled field formalism in order to determine the background field equations for the doubled space has been done in [36]. There it has been established that the background field equations arising from the one loop beta functional for the doubled formalism are the same as for the usual string. The doubled formalism can be seen as another description of string theory on target spaces described in the notation of [36] as locally $T^{n}$ bundles, with fibre coordinates $X^{i}$, over the base space $N$ with coordinates $Y^{a}$. The fibre of the bundle is then doubled to $T^{2n}$ with the coordinates denoted $\mathbb{X}^{A}$. The resulting sigma model Lagrangian will then be
\begin{equation}
\mathcal{L}=\frac{1}{4}\mathcal{H}(Y)d\mathbb{X}^{A}\wedge *d\mathbb{X}^{B}+\mathcal{L}(Y)+\mathcal{L}_{top}(\mathbb{X})
\end{equation}
where $\mathcal{L}(Y)$ is the string Lagrangian on the base, $\mathcal{H}(Y)$ is the metric on the fibre, and $\mathcal{L}_{top}$ is a purely topological term. 
Explicitly we have 
\[
  \mathcal{H}=
  \left( {\begin{array}{cc}
   h-bh^{-1}b & bh^{-1} \\
   -h^{-1}b & h^{-1} \\
  \end{array} } \right)
\]
where $h$ and $b$ are the target space metric and the $B$-field on the fibre of the undoubled space. We define $\mathbb{X}=(X^{i},\tilde{X}_{i})$ with $\{\tilde{X}_{i}\}$ being the coordinates on the T-dual torus. 
Reference [37] establishes the quantum equivalence of the doubled formalism to the usual sigma model formalism for worldsheets of arbitrary genus, provided the topological term mentioned above is added to the action. The topological term does not affect the classical theory but instead introduces relative signs in the sum over topological sectors. 
It depends only on the winding numbers around homology cycles in the worldsheet and hence does not affect the classical theory. The periodicities of the direct and dual coordinates are such that the $T^{d}$ torus parametrised by the dual coordinates is dual to the one parametrised by the direct coordinates. Then the term in the action $S_{top}=\frac{1}{2\pi\alpha'}\int\mathcal{L}_{top}$ becomes a sum of terms involving products of winding numbers for a conjugate pair of cycles, with a sum over 1-cycles. This leads to an alternating sign sum contribution to the functional integral given as a sum over winding numbers. This alternating sum involves the extra-dimensions arising due to the doubling and would not appear in an effective field theory. They will however influence the form of the corrections to the beta functions appearing in the low energy domain.

\section{Naturalness as a low energy effect of T-duality}
It is important to notice that the hierarchy problem is strongly related to the cosmological constant problem. It is known [15] that T-duality does not keep the cosmological constant, defined as the asymptotic value of the scalar curvature, invariant. First this has been noted in [16] where for an WZW model with a group $\tilde{SL(2,R)}$ a discrete subgroup has been gauged and relativity of the cosmological constant was first observed. Such a space has negative cosmological constant and under T-duality it is mapped into an asymptotically flat space. From a string theoretical perspective one can ask, together with the authors of [15] whether the standard low energy effective field theoretical definition of a cosmological constant makes sense. One can indeed wonder to what extent the cosmological constant can be seen as an unambiguous string observable. Ref. [15] notes that the change of the cosmological constant under T-duality is generic in the sense that it remains valid in higher dimensions as well. Starting with a ten dimensional type IIA theory with cosmological constant, a T-duality symmetry, if required to be a good symmetry, will force us to equate the cosmological constant to zero. Is it therefore possible that the network of string dualities will finally impose constraints on the cosmological constant, strong enough to explain its small value? 
A discussion of how making all scales of the matter sector functionals of the 4-volume element of the universe can remove the vacuum energy contributions from the field equations has been analysed in [24]. Such a discussion is compatible with large hierarchies between the Planck scale, electroweak scale, and vacuum curvature scale.

While the hierarchy problem and naturalness are related to the problem of the small cosmological constant, and may actually be one and the same problem, I will focus here on the naturalness issues. T-duality relates different geometries and/or topologies as seen from the perspective of extended objects (strings). As introducing such extended objects is equivalent to modifying a cohomology theory such that the coefficient structure replaces the normal geometric points with segments of algebraic curves, passing from a string to an effective description and back should be described simply in terms of modifying the coefficient structure of the cohomology theory in a controlled fashion. Duality symmetries are important in determining properties of the low energy effective theories and prove extremely useful in answering questions related to supersymmetry breaking.

To see how T-duality affects the low energy renormalisation group equations I consider a cutoff regularisation scheme where the relation between the bare and renormalised Green's function is 
\begin{equation}
\Gamma_{bare}(\{p^{2}\},1/\epsilon, \{g_{bare}\})=Z_{\Gamma}^{-1}(1/\epsilon, \{g_{\mu}\})\Gamma(\{p^{2}\},\mu,\{g_{\mu}\})
\end{equation}
where $g_{bare}=Z_{g}((1/\epsilon),\{g_{\mu}\})g$. Multiplication by the constant $Z_{g}$ obeys the group property and after eliminating the divergencies, such multiplications with finite constants are equivalent with making different choices of renormalisation schemes. Such schemes are therefore related by operations belonging to a Lie group. Given an arbitrary Green function $\Gamma$ obeying the normalisation condition $\Gamma(\{p^{2}\},\mu^{2},0)=1$ and looking for the variation of the Green's function with respect to the energy scale parameter $\mu$ one obtains 
\begin{equation}
\mu^{2}\frac{d}{d\mu^{2}}\Gamma=(\mu^{2}\frac{\partial}{\partial \mu^{2}}+\mu^{2}\frac{\partial g}{\partial \mu^{2}}\frac{\partial}{\partial g})\Gamma=\mu^{2}\frac{d ln Z_{\Gamma}}{d\mu^{2}}Z_{\Gamma}\Gamma_{bare}
\end{equation}
or, using the standard notation, we obtain the renormalisation group equation in partial derivatives
\begin{equation}
(\mu^{2}\frac{\partial}{\partial \mu^{2}}+\beta(g)\frac{\partial}{\partial g}+\gamma_{\Gamma})\Gamma(\{p^{2}\},\mu^{2},g_{\mu})=0
\end{equation}
The beta function and the anomaly dimensions are well known 
\begin{equation}
\begin{array}{l}
\beta=\mu^{2}\frac{dg}{d\mu^{2}}\rvert_{g_{bare}}\\
\\
\gamma_{\Gamma}=-\mu^{2}\frac{d ln Z_{\Gamma}}{d\mu^{2}}\rvert_{g_{bare}}\\
\end{array}
\end{equation}
By using characteristics we can write the general form of the solution of this equation as 
\begin{equation}
\Gamma(e^{t}\frac{\{p^{2}\}}{\mu^{2}},\bar{g}(t,g))e^{\int_{0}^{t}\gamma_{\Gamma}(\bar{g}(t,g))dt}
\end{equation}
where the characteristic equation is 
\begin{equation}
\begin{array}{ll}
\frac{d}{dt}\bar{g}(t,g)=\beta(\bar{g}), & \bar{g}(0,g)=g\\
\end{array}
\end{equation}
We call $\bar{g}(t,g)$ the effective coupling. One usually includes the vertex function and defines the whole product together as effective coupling 
\begin{equation}
g\Gamma(\frac{\{p^{2}\}}{\mu^{2}},g)
\end{equation}
If our Green function is an n-point function we can write the renormalisation of the coupling $g$ as $g_{bare}=Z_{\Gamma}Z_{2}^{-n/2}g$ and the the product is renormalised as $g\Gamma=Z_{2}^{n/2}g_{bare}\Gamma_{bare}$. One can construct a renormalisation group invariant quantity called the invariant charge $\xi$ by multiplying the product by the corresponding propagators 
\begin{equation}
\xi=g\Gamma(\frac{\{p^{2}\}}{\mu^{2}},g)\prod_{i}^{n}D^{1/2}(\frac{p_{i}^{2}}{\mu^{2}},g)
\end{equation}
The characteristic solution of the renormalisation group equation allows us to sum up an infinite series of logs coming from Feynman diagrams in both the IR $(t\rightarrow -\infty)$ or UV $(t\rightarrow \infty)$ regions and induces non-perturbative corrections to the otherwise perturbative expansion. As an example, the invariant charge in a massless theory with only one coupling constant is 
\begin{equation}
\xi(\frac{p^{2}}{\mu^{2}},g)=g(1+b\cdot g\cdot ln\frac{p^{2}}{\mu^{2}}+...)
\end{equation}
The beta function for a one loop approximation is $\beta(g)=b\cdot g^{2}$. The beta function can be written as a derivative of the invariant charge with respect to the log of the momentum 
\begin{equation}
\beta(g)=p^{2}\frac{d}{dp^{2}}\xi(\frac{p^{2}}{\mu^{2}},g)\rvert_{p^{2}=\mu^{2}}
\end{equation}
the RG-improved formula for the invariant charge is 
\begin{equation}
\xi_{RG}(\frac{p^{2}}{\mu},g)=\xi_{PT}(1,\bar{g}(\frac{p^{2}}{\mu^{2}},g))=\bar{g}(\frac{p^{2}}{\mu^{2}},g)
\end{equation}
where I replaced $t=ln\frac{p^{2}}{\mu^{2}}$. The effective coupling is a solution of the characteristic equation 
\begin{equation}
\begin{array}{lll}
\frac{d}{dt}\bar{g}(t,g)=b\bar{g}^{2}, & \bar{g}(0,g)=g, & t=ln\frac{p^{2}}{\mu^{2}}
\end{array}
\end{equation}
namely 
\begin{equation}
\bar{g}(t,g)=\frac{g}{1-b\cdot g\cdot t}
\end{equation}
When expanding this in terms of $t$ the geometrical progression above reproduces the expansion of the invariant charge with the difference that in this final expression, we also have an infinite series of terms $g^{n}t^{n}$ called the leading log approximation. For the next order in the logs one considers the next term in the expansion of the beta function and sums up the next series of terms in the form $g^{n}t^{n-1}$ and so on. The behaviour of the effective couplings is determined by the beta function. When the minimal subtraction scheme is employed, the renormalisation of the mass occurs in the same way as that of the couplings
\begin{equation}
m_{bare}=Z_{m}m
\end{equation}
The mass renormalisation constant $Z_{m}$ is independent of the mass parameters and depends only on dimensionless couplings. The effective mass is then given by 
\begin{equation}
\begin{array}{ll}
\frac{d}{dt}\bar{m}(t,g)=\bar{m}\gamma_{m}(\bar{g}), & \bar{m}(0,g)=m_{0}\\
\end{array}
\end{equation}
solving this together with the equations of the effective couplings one obtains 
\begin{equation}
\bar{m}(t,g)=m_{0}e^{\int_{0}^{t}\gamma_{m}(\bar{g}(t,g))dt}=m_{0}e^{\int_{g}^{\bar{g}}\frac{\gamma(g)}{\beta(g)}dg}
\end{equation}
Up to one loop one has $\beta(\alpha)=b\cdot \alpha^{2}$ and $\gamma_{m}(\alpha)=c\cdot \alpha$ with the solution 
\begin{equation}
m(t)=m_{0}(\frac{\alpha(t)}{\alpha_{0}})^{c/b}
\end{equation}
The mass is running due to radiative corrections. In the minimal subtraction scheme one may stop the running at $p^{2}=m^{2}$ and identify $m^{2}=\bar{m}^{2}(m^{2})$. A better way to define the mass is by identifying the physical mass with the pole mass. This does not depend on the scale and is scheme independent. The pole mass can be expressed by means of the running mass at the scale of a mass with finite and calculable corrections. If one calculates radiative corrections to the mass of the Higgs boson based entirely on the Standard Model, one obtains a loop integral of the form 
\begin{equation}
\int d^{4}p[\frac{1}{(\slashed{p}-m_{f})(\slashed{p}+\slashed{k}-m_{f})}]
\end{equation}
where $k$ is the Higgs momentum. This diverges quadratically for large $p$, independent of $k$ and hence generates a correction $\delta m^{2}\cong \Lambda^{2}$ where $\Lambda$ is the scale beyond which the low energy theory can no longer be applied. Let $\mu_{2}$ be the scale at which the breaking of the $SU(2)\times U(1)$ takes place. We assume that the standard model is the low energy theory of a more fundamental theory that rises at scale $\mu_{1}$. Using the fundamental theory one may imagine to be able to calculate the mass of the Higgs boson. The result of such a calculation would be a scale dependent mass parameter evaluated at the fundamental scale $\mu_{1}$. The important quantity for the low energy theory is the running mass calculated at the scale $\mu_{2}$. The relation between these two masses is given by 
\begin{equation}
m_{H}^{2}(\mu_{2})=m_{H}^{2}(\mu_{1})+C g^{2}\int_{\mu_{2}^{2}}^{\mu_{1}^{2}}dk^{2}+Rg^{2}+O(g^{4})
\end{equation}
where $g$ is a coupling constant, $C$ is dimensionless, and $R$ grows as a logarithm function with respect to $\mu_{1}$ as $\mu_{1}\rightarrow \infty$. The term proportional to $C$ diverges quadratically when $\mu_{1}\rightarrow \infty$. Usually, in order for $m^{2}_{H}(\mu_{2})\ll \mu_{1}^{2}$ one has to fine-tune the parameter $m_{H}^{2}(\mu_{1})$ to cancel the second term in the equation above which is of order $\mu_{1}^{2}$. The "natural" value for $m_{H}^{2}(\mu_{2})$ is a number of the order of $\mu_{1}^{2}$. Therefore what is the mechanism behind the fact that $\mu_{2}\ll \mu_{1}$? Given the experimental exclusion of various supersymmetric models at LHC, it appears implausible for supersymmetry to still be able to solve such a hierarchy problem. However, one may ask how the invariance of the high energy physical theory to topology changes induced by, say, T-duality may affect the low energy calculations given by the last equation? The requirement of topological invariance as demanded by T-duality indeed has some non-trivial effects on the renormalisation group equations. As said in the introduction, the beta functions are generators of the renormalisation group, while the group laws are given by the respective renormalisation group equations. To demand topological invariance of the theory with respect to high energy changes in topology amounts to modifications of the group laws which implicitly induce modifications in the form of the renormalisation group equations at low energies. Indeed, as will be shown further on, such modifications have precisely the form required to restore naturalness, i.e. they re-create terms similar to the terms obtained by adding supersymmetric partners, only this time, they originate simply from topological restrictions in the high energy end of the theory. The physical mass of the Higgs boson $m_{H}$ at tree level is proportional to the square root of the Higgs self-interaction coupling $\lambda$. We know that the observed value of $m_{H}$ is in the range that corresponds to vacuum metastability if there is no new physics between the electroweak scale and the Plank scale. Results for a full 2-loop calculation of the Higgs boson pole mass $m_{H}$ having the $\bar{MS}$ Lagrangian parameters $v, \lambda, y_{t}, g, g', g_{3}$ with the leading 3-loop corrections in the limit $(g_{3},y_{t})>>(\lambda, g, g')$ have been presented in [17]. In order to compute the Higgs boson physical mass $m_{H}$, the self energy function consisting of the sum of all 1-particle-irreducible 2-point Feynman diagrams needs to be calculated 
\begin{equation}
\Pi(s)=\frac{1}{16\pi^{2}}\Pi^{(1)}(s)+\frac{1}{(16\pi^{2})^{2}}\Pi^{(2)}(s)+...
\end{equation}
The complex pole squared mass is the solution of 
 \begin{widetext}
\begin{equation}
m_{H}^{2}-i\Gamma_{H}M_{H}=m^{2}_{B}+3\lambda_{B}v_{B}^{2}+\frac{1}{16\pi^{2}}\Pi^{(1)}(s_{pole})+\frac{1}{(16\pi^{2})^{2}}\Pi^{(2)}(s_{pole})
\end{equation}
\end{widetext}
In a Wilsonian RG the hierarchy problem appears as a radiative mixing of multiple relevant operators, caused by logarithmic divergences. Thus, because in Wilsonian RG, quadratic divergences determine the position of the critical surface in the theory space, and the scaling behavior of RG flows around the critical surface is determined only by the logarithmic divergences, we can subtract the quadratic divergences easily. The subtraction can therefore be interpreted as a choice of parametrization in the theory space [18].
The  required fine tuning appears therefore to be the distance of the bare parameters from the critical surface. This obviously means that quadratic divergences are not the main problem. However, Wilsonian RG radiative mixing implies that the lower mass scale is affected by higher scales through RG transformations. It is therefore important to understand what are the relevant high energy effects that can explain the non-naturalness. 
Consider the Higgs self coupling $\lambda$ introduced in the potential of $V^{(0)}=m^{2}H_{2}^{\dagger}H_{2}$ which we treat as a running low energy effective parameter. The RG improved Higgs mass can be written as 
\begin{equation}
(m_{H^{0}}^{2})_{RG}=\lambda v_{0}^{2}
\end{equation}
where the running quartic Higgs self coupling evaluated at the scale $s=m_{H^{0}}^{2}$ is 
\begin{equation}
\lambda = \int dt \beta_{\lambda}
\end{equation}
using the $\beta$ function at two-loop order $\beta_{\lambda}$ we obtain the next to leading log radiative corrections to the Higgs mass summed to all orders in perturbation theory [19]. This formula is defining for the self-coupling. However [20] showed that the beta function can be interpreted as a RG bulk space geometric curvature. When integrating over a curvature two form, the result may depend on the topological contributions to the beta function. 
Certain quantities constructed from the curvature on a differentiable principal bundle are inherent to the bundle and do not dependent on the specific curvature used for their definition. These are characteristic to the bundle, preserved by bundle diffeomorphisms and define topological invariants associated with the bundle. For details I refer the reader to [29] for further details on the construction of such invariants. 

In the most restricted case, the integral of the curvature can be interpreted as a topological invariant. An example is the generalisation of the Gauss-Bonnet theorem for higher dimensional spaces. In this case, denoting the curvature by $\Omega$ we may write $\int_{M}Pf(\Omega)=(2\pi)^{n}\chi(M)$ with $Pf(\Omega)$ the Pfaffian. For a four dimensional oriented manifold the Gauss-Bonnet theorem reads 
\begin{equation}
\frac{1}{32\pi^{2}}\int_{M}(|R_{m}|^{2}-4|R_{c}|^{2}+R^{2})d\mu=\chi(M)
\end{equation}
where $R_{m}$ is the Riemann curvature tensor, $R_{c}$ is the Ricci curvature tensor, and $R$ is the scalar curvature. In the general case we have the so called Chern-Weil theory which is capable of generating various other invariants. 
In the situation at hand, the beta function has been shown in [20] to be interpreted as a curvature and its integral is known to contribute to the running quartic Higgs self coupling. What is important is that a topological invariant arises in the form of the running quartic Higgs self coupling. This is particularly important because in order to impose topology invariance in agreement with the string geometric T-duality one needs to add a series of terms that would counter the effects of these topological invariants in order to make the theory independent of topology changes as induced by T-duality. Indeed, such terms can be added when the groups associated to different coefficient structures are equivalent as extensions and hence we cannot distinguish topological differences. In fact, this is precisely our requirement in agreement with the topology changing T-duality. 
Therefore, the renormalisation group equations will be changed due to high energy T-duality effects. These effects can be thought of as mixing between scales as visible in the Wilson renormalisation group approach. Particularly demanding that the renormalisation group operations are not sensitive to topological changes is equivalent to demanding that the low energy groups differ only by terms that do not distinguish the respective extensions to the analysed topologies. For group laws of the form $(m',g')(m,g)=(m'+m+\xi_{1}(g',g),g'g)$ the correction must be $\xi_{1}(g',g)=\xi_{2}(g',g)+\xi_{cob}(g',g)$. Translated into the RG equation language this provides us with topologically covariant terms that take into account the fact that the high energy theory must keep the universal coefficient theorem that leads to changes in the coefficients which would switch between detecting and not detecting topological features related by T-duality, trivial. Such shifts in the low energy theory will have strictly no physical effects (i.e. no particle detection whatsoever) because they are fundamentally string-geometrical effects that have no analogues for point-particle effective theories. However, they will have detectable effects on the radiative corrections to scalar masses due to the changes of the renormalisation group flows. Indeed the radiative corrections give a modification of the Higgs mass in the form of 
\begin{equation}
m^{*}=m^{2}-\frac{\lambda_{2}M^{2}}{32\pi^{2}}[1-\gamma_{E}+ln\frac{M^{2}}{4\pi\mu^{2}}]
\end{equation}
but demanding topological insensitivity at the string level leads, at low energies, according to the Verlinde duality [1], [2], to a redefinition of the renormalisation group laws namely to a modification of the renormalisation group equations. 
Considering that the renormalisation group equations are dual to equations of motion on the extra dimensions and the beta functions represent bulk space curvatures, after expressing the string theory in the bulk in a manifestly T-dual form, we double the string coordinates. Extra dimensional string coordinates correspond to generators of the renormalisation group and hence to beta functions. In modern language, this can be understood in terms of double field theory as resulting from incorporating string theoretical T-duality as a symmetry of a field theory defined on a double configuration space. T-duality however requires extended objects because it is based essentially on the existence of winding modes associated to wrappings of such extended objects around non-contractible cycles. A T-duality symmetric field theory will have to take the winding modes information into account. If we compactify the strings on a torus we obtain compact momentum modes, dual to compact coordinates $x_{i}$, $i=1,...,n$, as well as string winding modes. A new set of coordinates, $\tilde{x}_{i}$ dual to windings must also be considered for the compactified sector in the field theory. There is a vast literature on this field, including reviews like [35] on which I based my discussion up to now. In what concerns the low energy effective field theory, this doubling amounts to re-writing the RG equation as
\begin{equation}
(\mu\frac{\partial}{\partial \mu}+\beta_{\perp}(g)\frac{\partial}{\partial g}-\beta_{\parallel}(g)\frac{\partial}{\partial g}+\frac{1}{2}n\gamma(g))\Gamma^{(n)}(p_{i};g,\mu)=0
\end{equation}
where the beta function corresponding to the curvature has been split over the two terms, one over the original fibre bundle and one on the T-dual bundle used to T-symmetrise the bulk theory. These are of course only low energy remnants of the high energy T-duality symmetrised theory. Their existence in the low energy domain however implies cancellation of terms that would aid the naturalness of the effective theory. 
This can be rewritten by adding a term $\xi(g',g)$ which corrects the perturbative formulation of the beta function, a term corresponding to the one derived as the cobordism that restores topological insensitivity by demanding that the two coefficient structures in cohomology generate extensions within the universal coefficient theorem that do not single out any topological change.  Here $g'$ and $g$ are nothing but the parameters defined at the two different scales. I rewrite therefore, heuristically 
\begin{equation}
(\mu\frac{\partial}{\partial \mu}+(\beta(g)+\xi(m,M))\frac{\partial}{\partial g}+\frac{1}{2}n\gamma(g))\Gamma^{(n)}(p_{i};g,\mu)=0
\end{equation}

Interestingly enough, while keeping the prescription that the modifications brought to the low energy effective theory must obey the laws of topological insensitivity imposed by the universal coefficient theorem's extensions, such terms induce additional scale effects that, because of their dependence on the topology at a specific order (considering for example the large N topological expansion) will contribute with different signs. For example torus terms will annihilate sphere terms, due to the requirement of T-duality, etc. Therefore we have

\begin{equation}
m^{*}=m^{2}-\frac{\lambda_{2}M^{2}}{32\pi^{2}}[1-\gamma_{E}+ln\frac{M^{2}}{4\pi\mu^{2}}+\sum_{i}(-1)^{i}\zeta(m^{2},M^{2})]
\end{equation}
which will introduce the corrections to the extreme scale dependence. The Wilsonian physical mixing effect is accounted for by the scale mixing between stringy effects and effective field theoretical effects.

\section{Conclusions}
This article provides a heuristic solution to the hierarchy problem by explaining it in terms of a non-trivial mixing of UV and IR effects. In modern terminology, we may understand the naturalness of the standard model hierarchy as a dual phenomenon to the existence of high energy T-duality and particularly of bulk-space invariance to topology changes when explicit T-duality invariant string theory is employed. 
As there is no direct reason for such a requirement in the low energy domain where objects are point-like it comes as no surprise that the hierarchy problem remained so long with no clear resolution.

\end{document}